\documentclass{ws-p10x7}

\begin{document}

\title{Anisotropic color superconductor}

\author{J. Ho\v{s}ek}

\address{Dept. Theoretical Physics, Nuclear Physics Institute,25068 \v{R}e\v{z} (Prague), 
Czech Republic\\E-mail: hosek@ujf.cas.cz}

\twocolumn[\maketitle\abstract{
We argue that the QCD matter not far above a critical confinement-deconfinement baryon density 
and low temperatures can develop spontaneously the condensates of spin-one quark Cooper
pairs. Depending upon their color these condensates characterize two distinct anisotropic 
color-superconducting phases. For them we derive the generic form of the quasiquark 
dispersion laws and the gap equation. We also visualize the soft Nambu-Goldstone modes of
spontaneously broken global symmetries, and demonstrate an unusual form of the Meissner effect.}]

\section{Basic picture}%1

With QCD as the microscopic theory of strong interactions it became mandatory to pursue,
if possible, both experimentally and theoretically, all corners of its phase diagram\cite{wilczek}.
Here we restrict our attention to that of high baryon densities and very low temperatures:
Not far above a critical confinement-deconfinement baryon density $n_c \sim 5 n_{nucl.matter}\sim
 0.72/fm^3$ and at low $T$ the deconfined QCD matter should be a rather strongly interacting quantum
many-colored-quark system. Its detailed actual behavior in the considered region depends 
solely upon the details of the effective interactions relevant there.

For definiteness (and because we think it is both natural and simplifying) we assume 
that the strong (but nonconfining) gluon interactions dress the tiny quark masses $m_u, m_d$
(we restrict our discussion to the case of two light flavors) into a common larger effective
mass $m_*$, and become weak. Residual interaction between the massive (quasi)quark 
excitations $\psi^{a}_{\alpha A}$ ($a$ - color, $\alpha$ - Dirac, $A$ - flavor $SU(2)$ indices)
can then be described by appropriate short-range (approximately contact) four-fermion interactions
${\cal L}_{int}$. Some pieces of ${\cal L}_{int}$ even have a solid theoretical justification:
the instanton-mediated interaction of t'Hooft\cite{thooft}, and the Debye-screened chromoelectric
one-gluon exchange. For our case of fragile ordered phases to be discussed below
one should keep in mind also yet unknown effective local four-fermion interactions due to the 
exchanges of heavy collective excitations eventually existing in more robust phases.
The resulting effective Lagrangian
\begin{eqnarray}
 {\cal L}_{eff} &= & \overline \psi(i\gamma^{\mu}D_{\mu} - m_{*} + \mu \gamma_{0})\psi
 \nonumber \\
&&{} - \frac{1}{4}F_{a\mu\nu}F^{a\mu\nu} +{\cal L}_{int}
\label{eq:effective}
\end{eqnarray}
in which the gluon interactions are treated perturbatively (and neglected in the lowest approximation)
defines a relativistic version of the Landau Fermi-liquid concept. The quark chemical 
potential of interest is of the order of $\mu \sim 500 MeV$, whereas $m_*$ is to be 
determined experimentally.

By assumption, ${\cal L}_{eff}$ is exactly $SU(3)_c \times SU(2)_{I} \times U(1)_{V} \times
O(3)$ invariant. There is no approximate $SU(2)$ chiral symmetry of (\ref{eq:effective})
which could be broken spontaneously. Hence, there should be no Nambu-Goldstone(NG)
pions in dense and cold deconfined phase(s) of QCD. It would be also misleading to think of 
$\psi$ and $m_*$ as of the constituent quark and of the constituent mass. There is nothing
they might constitute.

At present, there are no experimental data, either real or the lattice ones which would
check our assumption. For our considerations it is not, however, essential. An alternative
picture, and in fact the more commonly discussed one is that in the cold deconfined QCD
matter the $u, d$ quarks stay approximately massless at the Lagrangian level
as they were in the confined phase. Discussion of the superfluid\footnote{We do not
sharply distinguish between superconductivity and superfluidity 
in many-fermion systems. The superconductor is a superfluid having the perturbative
long-range gauge interactions switched on. Superfluidity in many-boson systems
(for example in $^{4}He$) is, however, an entirely different story.} phases presented below 
applies also to this case. On top of that it is, however, obligatory to ask
(and to answer) how the (approximate) chiral symmetry is realized in this case. 

The cold and dense deconfined QCD matter should exist in the interiors of the neutron stars,
and optimistically also in the early stages of the relativistic heavy-ion collisions
studied experimentally with much effort at present. Consequently, theoretical studies of such a matter 
are more than an intellectual challenge.

\section{Isotropic superconductors}%2

In principle, the behavior of a cold and dense deconfined QCD matter governed by (\ref{eq:effective})
should be similar to that of any non-relativistic low-$T$ Landau Fermi-liquid (say of
electrons in metals or of the atoms of liquid $^{3}He$). The differences are rather technical: 
(1)Characteristic energies given by the chemical potential $\mu \sim O(500 MeV)$ 
require the relativistic description. (2)The quarks carry, besides spin, also the 
flavor and color. (3)The gauge fields are both Abelian (photon) and non-Abelian (gluons).
(4) The origin of the effective interactions is different.

In fact, the behavior of any (non-relativistic and relativistic) quantum many-fermion liquid 
of the Landau type is uniquely dictated by "theorems": (1)When scaling the fermion momenta
towards the Fermi surface all interactions but one become irrelevant\cite{pol,evans}.
This implies that almost all such systems should behave thermodynamically as a corresponding noninteracting
system of fermions. For example, the specific heat should grow linearly with $T$. Such a behavior
is indeed observed in the low-$T$ electron systems in metals, and in the liquid $^{3}He$.
We are not aware of any experimental data in the relativistic systems. (2)The four-fermion 
interaction attracting fermions with opposite momenta at the Fermi surface is the only exception: 
Even if arbitrarily small, it causes the (Cooper) instability of the filled Fermi sea with respect
to spontaneous condensation of the fermion Cooper pairs with opposite momenta into a more
energetically favorable ground state. The new ground state, being by construction and by definition 
translationally invariant, has in the simplest nonrelativistic case of the ordinary local BCS-type four-
fermion interaction the property
\begin{equation}  
\langle \psi^{+}_{\alpha}(x)(\sigma_{2} )_{\alpha \beta} \psi^{+}_{\beta}(x) \rangle = \Delta \neq 0
\label{eq:psipsi}
\end{equation}
It clearly exhibits spontaneous breakdown of the $U(1)$ phase symmetry generated 
by the operator of the particle number. Given a four-fermion interaction, the BCS theory 
provides for the microscopic quantitative understanding of the system.

Potential relevance of the physics of superconductors for the cold and dense QCD matter 
was recognized long ago\cite{collins,frautschi,bailin}. Recent abrupt 
increase of interest in this idea was driven by two influential papers\cite{arw,shuryak}:
By explicit calculations they found that the realistic four-fermion interaction of the t'Hooft type
\cite{thooft} gives rise to the color-superconducting isotropic (spin-zero) phase I
characterized by the condensate
\begin{equation}
\langle \overline {\psi}_{\alpha aA}(x) {\epsilon}^{ab3} (\tau_2)_{AB} (\gamma_5C)_{\alpha \beta}
\overline {\psi}_{\beta bB}(x)\rangle = \Delta
\label{eq:phase1}
\end{equation}  
which is phenomenologically quite appealing: $\Delta \sim 100MeV$. The condensate (\ref{eq:phase1})
breaks spontaneously the $SU(3)_{c} \times U(1)$ symmetry down to $SU(2)_{c}$.
Consequently, when the gauge interactions are switched off, there are 5+1 NG gapless 
collective excitations in the spectrum. Clearly, they can be only exited
by the quark bilinears. When the gauge interactions are switched on
5 gluons acquire masses by the underlying Higgs mechanism\cite{rischke}.
Recently,properties of this phase (and of its three-flavor relative) were studied in great detail.  

Pauli principle allows for yet another isotropic, color-superconducting phase II. It is characterized 
by the condensate (we introduce $T_{3} = \tau_{3} \tau_{2}$)
\begin{equation}
\langle \overline {\psi}_{\alpha aA}(x)(T_{3})_{AB} (\gamma_5C)_{\alpha \beta}
\overline {\psi}_{\beta bB}(x)\rangle = \Delta_a \delta_{ab}
\label{eq:phase2}
\end{equation}
While $\Delta$ in (\ref{eq:phase1}) corresponds to the vacuum expectation value (vev) of 
the spin-0, color triplet, isospin-0 Higgs field, the condensate (\ref{eq:phase2})
corresponds to the vev of the spin-0, color sextet, isospin-1 Higgs field. The phase II
is theoretically interesting mainly by spontaneous breakdown of truly global isospin symmetry.
Since there is nobody who might "eat" them, the two corresponding NG gapless excitations remain in the 
physical spectrum, and become thermodynamically important. Generic form of the excitations
in this phase is the same as in the phase I. Different is merely their counting.

\section{Anisotropic superconductors}%3

The Pauli principle itself allows for yet another two condensates,
both having spin one\cite{hosek}: (1)The anisotropic color-superconducting phase III is characterized by
\begin{equation}
\langle \overline {\psi}_{\alpha aA}(x)(\tau_2)_{AB}(\gamma_0\gamma_3C)_{\alpha \beta}
\overline{\psi}_{\beta bB}(x)\rangle = \Delta_{a}\delta_{ab}
\label{phase3}
\end{equation}
which corresponds to the vev of isospin-0, color sextet antisymmetric tensor Higgs field 
$\Phi_{ab \alpha \beta}(x)$ describing spin-1. (2)The phase IV is characterized 
by the condensate
\begin{equation}
\langle \overline{\psi}_{\alpha aA}(x)(T_{3})_{AB} \epsilon^{ab3}(\gamma_0\gamma_3C)_{\alpha \beta}
\overline {\psi}_{\beta bB}(x) \rangle = \Delta
\label{phase4}
\end{equation}
which corresponds to the vev of isospin-1, color triplet, antisymmetric tensor Higgs field
$\Phi^{c}_{I \alpha \beta}(x)$. 

Clearly, only the detailed behavior of ${\cal L}_{int}$ 
can select which one out of the four physically distinct ordered phases is the most 
energetically favorable one.

The anisotropic phases are interesting for they break down spontaneously the rotational symmetry like
ferromagnets. In relativistic systems this is certainly not a very frequent phenomenon. It is possible 
only at finite quark density which itself breaks down explicitly the Lorentz invariance. 
The translational invariance of the ground state must of course remain inviolable.

In order to find out the generic features of the excitations of anisotropic superconductors
we have analyzed a model having all necessary properties but less indices. It is defined by its
Lagrangian
\begin{eqnarray}
 {\cal L}_{eff} &=& \overline \psi(i\gamma^{\mu}D_{\mu} - m_{*} + \mu \gamma_{0})\psi
 \nonumber\\
 &&{} - \frac{1}{4}F_{\mu\nu}F^{\mu\nu}+{\cal L}_{int}
\label{eq:abel}
\end{eqnarray} 
where 
\begin{equation}
{\cal L}_{int} = G [(\overline \psi \gamma_{0} \psi)^2 - (\overline \psi \gamma_{0}
\vec \tau \psi)^2]
\label{eq:abelint}
\end{equation}
The Lagrangian (\ref{eq:abel}) is $U(1)$ gauge invariant, and $SU(2)_{V}\times O(3)$
globally invariant. The interaction (\ref{eq:abelint}) is chosen in such a way that the 
isotropic (spin-0) condensate 
\begin{equation}
\langle \overline \psi_{\alpha A}(x)(T_{3})_{AB} (\gamma_{5}C)_{\alpha \beta} \overline \psi_{\beta B}(x)\rangle 
\nonumber\\
\end{equation}
identically vanishes. Thus, either (\ref{eq:abel}) can be treated perturbatively, or the 
interaction (\ref{eq:abelint}) gives rise to the anisotropic condensate
\begin{equation}
\langle \overline \psi_{\alpha A}(x)(\tau_{2})_{AB} (\gamma_{0} \gamma_{3} C)_{\alpha \beta}
\overline \psi_{\beta B}(x) \rangle = \Delta
\label{eq:abelcond}
\end{equation}

Introducing the field ($\psi^{\cal C} = C \overline \psi$)
\begin{equation}
q_{\alpha A} = \frac{1}{\sqrt{2}}
\left(\begin{array}{c}\psi_{\alpha A}\\ [0.1in]
(\tau_{2})_{AB} \psi^{\cal C}_{\alpha B}
\end{array}\right)
\label{eq:q}
\end{equation}
we define a new self-consistent perturbation theory by the bilinear Lagrangian
\begin{equation}
{\cal L}^{'}_{0} = \overline q S^{-1}(p)q = \overline q S^{-1}_{0}(p)q - {\cal L}_{\Delta}
\nonumber\\
\end{equation}
and the new interaction ${\cal L}^{'}_{int} = {\cal L}_{int} + {\cal L}_{\Delta}$ where
\begin{equation}
{\cal L}_{\Delta} = \overline q
\left(\begin{array}{cc} 0& \gamma_{0}\gamma_{3}\Delta\\ [0.1in]
\gamma_{0}(\gamma_{0}\gamma_{3}\Delta)^{+}\gamma_{0}& 0
\end{array}\right)q
\nonumber\\
\end{equation}
Finding $S(p)$ amounts to finding the form of the quasiquark dispersion laws. They have the form 
\begin{equation}
E_{(1)}(\vec p)= \left(\epsilon^2_{p} + |\Delta|^2 + \mu^2 + D^{2}(\vec p)\right)^{1/2}
\label{eq:e1}
\end{equation}
\begin{equation}
E_{(2)}(\vec p)= \left(\epsilon^2_{p} + |\Delta|^2 + \mu^2 - D^{2}(\vec p)\right)^{1/2}
\label{eq:e2}
\end{equation}
where $\epsilon_{p} = \sqrt{\vec p^{2} + m_{*}^2}$, and
\begin{equation} 
D^{2}(\vec p)=2(\epsilon^{2}_{p}\mu^{2} + (p_{1}^{2}+p_{2}^{2}+m_{*}^{2})|\Delta|^{2})^{1/2}.
\nonumber\\
\end{equation}
The equations (\ref{eq:e1}) and (\ref{eq:e2})
explicitly demonstrate spontaneous breakdown of the rotational symmetry of (\ref{eq:abel}).  

Requirement that ${\cal L}_{int}^{'}$ gives zero contribution 
to $S^{-1}(p)$ in the lowest self-consistent
approximation results in the equation for the gap $\Delta$:
\begin{eqnarray}
&&\Delta + 2\Delta G \int \frac{d^3p}{(2\pi)^3} \left( \frac{1}{E_{(1)}(\vec p)} +
\frac{1}{E_{(2)}(\vec p)} \right)\nonumber\\
&&\left( 1 - \frac{4(p_1^2 + p_2^2)}{E_{(1)}(\vec p)E_{(2)}(\vec p)}\right ) = 0
\label{eq:gap}
\end{eqnarray}
Numerical analysis of Eq.(\ref{eq:gap}) yet remains to be done. 
Here we simply assume that 
for the properly regularized integral\cite{arw} in (\ref{eq:gap})$\Delta \neq 0$ does exist. 
It is interesting to note 
that $E_{(2)}$ vanishes at $p_1^2 + p_2^2 = |\Delta|^2 + \mu^2 - m_{*}^2, p_3 = 0$. Fortunately
for the gap equation the circle of vanishing $E_{(2)}$ does not lie on the Fermi surface 
where the interaction is relevant.

For the gauge interaction switched off the condensate (\ref{eq:abelcond}) breaks spontaneously the global 
$U(1) \times SU(2) \times O(3)$ symmetry down to $SU(2) \times O(2)$. Consequently, 
there should exist $1+2$
NG quark-composite excitations. Their quantum numbers are found by analyzing the Goldstone commutator.
In terms of the field $q$ and of the Pauli matrices $\Gamma_{i}$ which operate in the space
of $q$ the NG composites have the form $\overline q\gamma_{0} \gamma_{3}(\Gamma_{1}-i\Gamma_{2})q,
i\overline q\gamma_{0} \gamma_{2}(\Gamma_{1}+i\Gamma_{2})q, i\overline q\gamma_{0}\gamma_{1}
(\Gamma_{1}+i\Gamma_{2})q$.

Finally, it would be desirable to know how the anisotropic condensate (\ref{eq:abelcond})
influences the behavior of the gauge field when perturbatively switched on. We plan to 
address this question within the microscopic description (\ref{eq:abel}) in a future work.
Here we present a straightforward analysis within the Higgs approach taking for granted 
that the condensate (\ref{eq:abelcond}) is a vev of the antisymmetric order parameter 
$\Phi^{\mu \nu}$ describing the spin one. Requirement that only its time-space components propagate 
fixes the form of its kinetic term which has to be gauged. The self-interaction of the 
field $\Phi^{\mu \nu}$ is chosen in such a way that $\langle \Phi_{03}\rangle = \Delta$.
The resulting effective Lagrangian has the form
\begin{equation}
{\cal L}_{H} = - (D^{\lambda}\Phi_{\lambda \mu})^{+} D_{\nu}\Phi^{\nu \mu} - V(\Phi)
\label{eq:higgs}
\end{equation}
in which we have ignored for simplicity the fact that (\ref{eq:higgs}) should be only $O(3)$
invariant. The mass term of the gauge field following from (\ref{eq:higgs}) has the form
\begin{eqnarray}
{\cal L}_{mass}&=&e^2|\Delta|^2(A_{0}^2-A_{3}^2)= e^2|\Delta|^2 A_{\mu}A^{\mu}\nonumber\\
&&  + e^2|\Delta|^2(A_{1}^2 + A_{2}^2)
\label{eq:meissner}
\end{eqnarray}
We think that the anisotropic Meissner-Higgs effect (\ref{eq:meissner}) is peculiar but not 
unexpected. The first term $e^2|\Delta|^2 A_{\mu}A^{\mu}$ is the effect independent of the spin
of the order parameter, whereas the second term represents an anisotropic "anti-Meissner-Higgs" effect
due to the spin. 

\section{Conclusion}%4

It is gratifying to observe that QCD has a corner in its phase diagram which is accessible experimentally,
and which should be full of new phenomena associated with macroscopic quantum ordered  
phases of the superfluid type.

Although the analogy between superconductivity and the low-$T$ deconfined QCD matter seems 
rather strong, it falters in one important respect: For studying the color-superconducting
phases we can safely forget about using external chromomagnetic and external chromoelectric
fields. We believe that due to the macroscopic quantum nature of these phases it is nevertheless 
justified to speculate that their experimental signatures might be brighter than
those of the 'ordinary' quark-gluon plasma above a superconducting $T_{c}$.

 \section*{Acknowledgments}

Part of the work was done during the program "QCD at Finite Baryon Density"
organized by the INT in Seattle. I am grateful to the Institute for the financial support,
and to Mike Alford, Michael Buballa, Krishna Rajagopal, Edward Shuryak, Jac Verbaarschot,
and Uwe Wiese for many discussions. The work was also supported by
the Committee for the CERN-CR Cooperation, and by the grant GACR 202/0506.

\end{document}